\documentclass[twocolumn,showpacs,preprintnumbers,amsmath,amssymb,superscriptaddress]{revtex4}
\usepackage{graphicx}
\usepackage{braket}
\usepackage{here}
\begin{document}
\title{Observation of Devil's Staircase in the Novel Spin Valve System SrCo$_6$O$_{11}$}
\author{T.~Matsuda}
\affiliation{Department of Applied Physics and Quantum-Phase Electronics Center (QPEC), 
University of Tokyo, Hongo, Tokyo 113-8656, Japan}
\author{S.~Partzsch}
\affiliation{Leibniz Institute for Solid State and Materials Research IFW Dresden, 
Helmholtzstrasse 20, 01069 Dresden, Germany}
\author{T.~Tsuyama}
\affiliation{Department of Applied Physics and Quantum-Phase Electronics Center (QPEC), 
University of Tokyo, Hongo, Tokyo 113-8656, Japan}
\affiliation{Institute for Solid State Physics, University of Tokyo, 
Kashiwanoha 5-1-5, Chiba 277-8581, Japan}
\author{E.~Schierle}
\affiliation{Helmholtz-Zentrum Berlin f\"ur Materialien und Energie, 
Albert-Einstein-Str.15, 12489 Berlin, Germany}
\author{E.~Weschke}
\affiliation{Helmholtz-Zentrum Berlin f\"ur Materialien und Energie, 
Albert-Einstein-Str.15, 12489 Berlin, Germany}
\author{J.~Geck}
\affiliation{Leibniz Institute for Solid State and Materials Research IFW Dresden, 
Helmholtzstrasse 20, 01069 Dresden, Germany}
\author{T. Saito}
\affiliation{Institute for Chemical Research, 
Kyoto University, Uji, Kyoto 611-0011, Japan} 
\author{S.~Ishiwata}
\affiliation{Department of Applied Physics and Quantum-Phase Electronics Center (QPEC), 
University of Tokyo, Hongo, Tokyo 113-8656, Japan}
\author{Y.~Tokura}
\affiliation{Department of Applied Physics and Quantum-Phase Electronics Center (QPEC), 
University of Tokyo, Hongo, Tokyo 113-8656, Japan}
\affiliation{RIKEN Center for Emergent Matter Science (CEMS), Wako, 
Saitama 351-0198, Japan}
\author{H.~Wadati}
\email{wadati@issp.u-tokyo.ac.jp}
\homepage{http://www.geocities.jp/qxbqd097/index2.htm}
\affiliation{Department of Applied Physics and Quantum-Phase Electronics Center (QPEC), 
University of Tokyo, Hongo, Tokyo 113-8656, Japan}
\affiliation{Institute for Solid State Physics, University of Tokyo, 
Kashiwanoha 5-1-5, Chiba 277-8581, Japan}
\pacs{71.30.+h, 71.28.+d, 73.61.-r, 79.60.Dp}
\date{\today}
\begin{abstract}
Using resonant soft x-ray scattering as a function of 
both temperature and magnetic field, we reveal a large 
number of almost degenerate magnetic orders in SrCo$_6$O$_{11}$. 
The Ising-like spins in this frustrated material in fact 
exhibit a so-called magnetic devil's 
staircase. It is 
demonstrated how a magnetic field induces transitions 
between different microscopic spin configurations, 
which is responsible for the magnetoresistance of SrCo$_6$O$_{11}$. 
This material therefore constitutes a
unique combination of a magnetic devil's 
staircase and spin 
valve effects, yielding a novel type of magnetoresistance system. 
\end{abstract}
\pacs{71.30.+h, 71.28.+d, 79.60.Dp, 73.61.-r}
\maketitle
Combining different materials in artificial nanostructures is 
a most important approach to create improved or even completely 
new electronic functionalities for technological applications. 
A very prominent example for this is the giant magnetoresistance 
(GMR), which was first realized by multilayers of 
alternating nonmagnetic and ferromagnetic metals \cite{nobel1,nobel2} and 
which now is an indispensable part of today's information technology. 
In these GMR systems the electrical resistance is high 
for an antiparallel alignment of the magnetization 
in the neighboring magnetic layers, 
while it is low for a parallel alignment of those magnetizations. 
For this reason such systems are also referred to as spin valves. 

\begin{figure}
\begin{center}
\includegraphics[width=9cm]{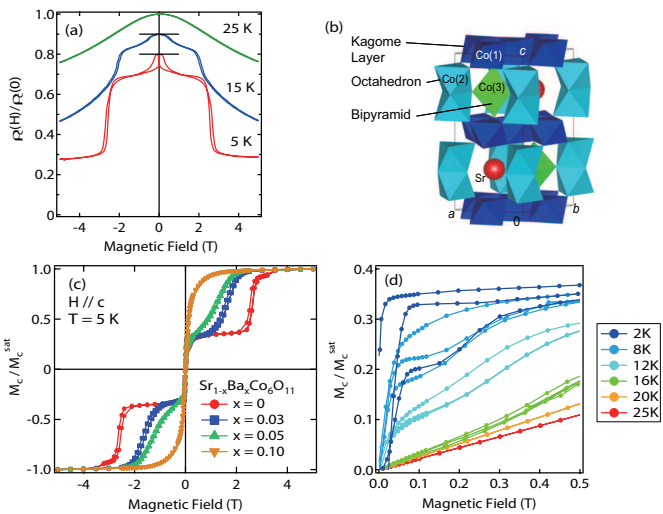}
\caption{(Color online) 
(a) Normalized out-of-plane resistivity $\rho_c(H)/\rho_c(0)$ 
of SrCo$_6$O$_{11}$ ($H$//$c$) taken from Ref.~\cite{ishiwataPRL}.
(b) Crystal structure of SrCo$_6$O$_{11}$. 
(c) Out-of-plane magnetization of Sr$_{1-x}$Ba$_x$Co$_6$O$_{11}$ 
as a function of magnetic field ($H$//$c$) at 5 K. 
(d) Temperature dependence of out-of-plane magnetization 
of SrCo$_6$O$_{11}$ in low magnetic field region ($H$//$c$).}
\label{fig1}
\end{center}
\end{figure}

Large or even colossal magnetoresistance can also occur as 
an intrinsic effect in bulk materials, 
due to the interplay of mobile charge carriers 
and localized spins. 
Here the doped manganites provide the probably 
most famous examples \cite{RamirezMn,RaoMn,PrellierMn,Hungry,TokuraMn}. 
A particularly interesting material with intrinsic magnetoresistance 
is the recently discovered Co oxide SrCo$_{6}$O$_{11}$, 
the lattice structure of which is believed to realize a GMR multilayer
system at the atomic scale \cite{synthesis}. Figure \ref{fig1} (a) 
shows the normalized out-of-plane resistivity $\rho_c(H)/\rho_c(0)$ 
of SrCo$_6$O$_{11}$ ($H$//$c$), which is a clear manifestation 
of magnetoresistance in this material \cite{ishiwataPRL}. 
As shown in Fig.~\ref{fig1} (b), SrCo$_6$O$_{11}$ exhibits a layered 
crystal structure consisting of three parts: 
(i) metallic Kagome-layers formed by the edge-sharing octahedra, 
(ii) dimerized octahedra and 
(iii) trigonal bipyramids. 
In the following the Co-sites of the Kagome-layers, 
the dimerized octahedra and trigonal bipyramids are referred 
to as Co(1), Co(2) and Co(3), respectively 
(cf. Fig.~\ref{fig1} (b)). 
It was shown earlier that the magnetism of SrCo$_{6}$O$_{11}$ 
is due to localized Ising-like spins of the 
Co(3)-sites, whereas the charge transport 
happens in the subsystems containing Co(1) and Co(2) \cite{ishiwataPRL}. 
The metallic Kagome-layers are hence linked by magnetic layers 
containing Co(3) and therefore realize a GMR-multilayer structure at the atomic level. 
The observed strong Ising type anisotropy along the $c$ axis 
can be explained by a non-vanishing orbital moment and 
the resulting spin-orbit coupling of Co(3), 
as discussed earlier for Ca$_3$Co$_2$O$_6$, 
which also contains Co-sites with trigonal local symmetry \cite{CCO}.  

One of the most striking magnetic features of SrCo$_6$O$_{11}$ 
observed so far are plateaus in the magnetization as 
a function of the applied magnetic field 
along the $c$-axis \cite{synthesis} as shown in Fig.~\ref{fig1} (c). 
These plateaus correspond to 1/3 and 3/3 of the saturated 
moment \cite{synthesis} and were found 
to reflect different stackings along $c$, 
namely an up-up-up structure 
for the 3/3 phase and an up-up-down configuration 
for the 1/3 phase \cite{SaitoJMMM}. 
As in the case of an artificial 
GMR-multilayer, the transition between these magnetic phases are 
thought to cause the giant magnetoresistance of this compound. 
Specifically, the magnetic up-up-up structure of 
the 3/3 phase \cite{ishiwataPRL} leads to less spin scattering and 
therefore a smaller electrical resistivity than in the up-up-down 
configuration of the 1/3 phase \cite{ishiwataPRL}. 

In this letter, we present a high-resolution 
resonant soft x-ray scattering (RSXS) 
study of SrCo$_6$O$_{11}$ 
bulk single crystals as a function of temperature ($T$) 
and applied magnetic field ($H$). 
We discover various metastable magnetic orders 
in low magnetic fields that escaped detection in earlier experiments. 
The occurrence of all these metastable phases can be 
interpreted in terms of strong frustration 
in a strongly anisotropic magnetic system, 
which results in a large variety of almost 
degenerate magnetic structures 
\cite{Bak,JRM1,JRM2,Ohwada,JvB,Bak2,Selke,Nakanishi}. 
In fact, our results imply that SrCo$_6$O$_{11}$ is the first realization of 
a magnetic devil's 
staircase in a $3d$-electron system. 
Our central result therefore is that magnetic frustration 
is a fundamental ingredient for the functionality of 
SrCo$_6$O$_{11}$ which realizes a novel and sensitive 
GMR-system at the atomic level in a single phase material. 
In addition, the high degree of frustration intrinsically 
causes a strong sensitivity of the system 
to external modifications which consequently 
opens the possibility of tailoring functionality 
as demonstrated in this study by Ba substitution 
at the Sr site. 

\begin{figure}
\begin{center}
\includegraphics[width=9cm]{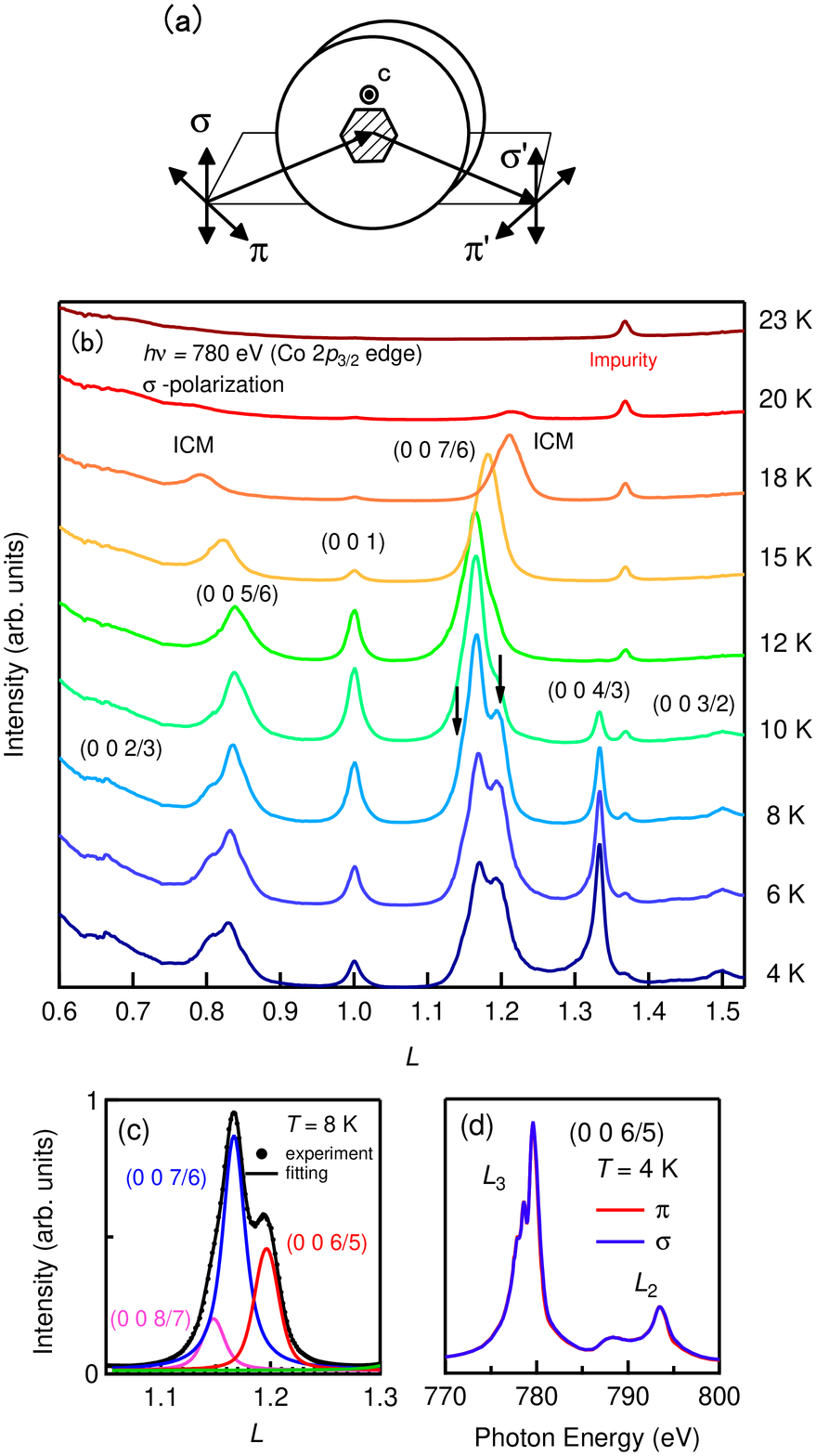}
\caption{(Color online) 
(a) Experimental geometry for RSXS measurements. 
The arrows indicate the directions of polarizations of 
x-rays. 
(b) Magnetic peak profile of SrCo$_6$O$_{11}$ 
at various temperatures 
at zero magnetic field. 
(c) Magnetic peak fitting of SrCo$_6$O$_{11}$ 
around $L$ = 7/6. 
(d) Photon-energy dependence of intensity of the 
$L$ = 6/5 reflection for $\sigma$ and $\pi$ 
polarizations.}
\label{peak}
\end{center}
\end{figure}

Pure and Ba-substituted SrCo$_6$O$_{11}$ bulk single 
crystals were synthesized 
by the high-pressure technique \cite{synthesis}. 
The typical sample size was 
$\simeq 0.20 \times 0.20 \times 0.05$ mm$^3$. 
The out-of-plane magnetization of 
pure and Ba-substituted SrCo$_6$O$_{11}$ are
shown in Figs.~\ref{fig1} (c) and (d). 
RSXS is a powerful tool to reveal ordered structures 
in solids such as magnetic, charge and orbital ordering 
\cite{RSXDreview,Wadati,Partzsch,Zhou,Beale,Smadici}. 
Here we employed the strongly enhanced magnetic 
sensitivity of RSXS at the  Co $2p_{3/2}$ edge 
(780 eV) in order to investigate the subtle magnetic phase 
transitions in our small-volume samples 
depending on temperature and external magnetic fields. 
The present study is one of the very first 
RSXS experiments performed under magnetic fields 
of several Tesla. 
The experiments were carried out 
at the High-field Diffractometer operated 
at the UE46-PGM1 beamline of BESSY-II, Germany. 
Figure \ref{peak} (a) shows the scattering 
geometry used for the present experiments. 
Temperatures down to 4 K could be reached 
using a continuous helium-flow cryostat. 
X-ray polarization was linear ($\sigma$ and $\pi$). 
Applied magnetic field was up to 4 T. 

Figure \ref{peak} (b) shows the diffraction peaks 
of SrCo$_6$O$_{11}$ 
at zero field for various temperatures. 
Quite surprisingly and very uncommon for RSXS experiments, 
a large number of superlattice reflections 
at $L=$ 2/3, 5/6, 1, 7/6, 4/3 and 3/2 is observed. 
The small and temperature independent peak at $L=1.37$ 
is assigned to some impurity in the sample 
because it does not show temperature dependence. 
$L$ = 1 commensurate (CM) peak and two incommensurate (ICM) 
peaks around $L$ = 0.8, 1.2 appear at 20 K ($T_{c1}$). 
These ICM peaks move to $L$ = 5/6 and 7/6, respectively, 
as the temperature is decreased, and finally are locked 
at these values at 12 K ($T_{c2}$), respectively. 
At $T_{c2}$, there appear 
$L$ = 5/6 and 8/7 shoulders of the $L$ = 7/6, and 
simultaneously $L$ = 2/3, 4/3 and 3/2 peaks. 

Intensities of all the magnetic peaks were independent 
of the polarizations $\sigma$ and $\pi$, as shown 
in Fig.~\ref{peak} (d) for the case of $L=6/5$. 
For a trigonal local symmetry 
and spins along the $c$ direction, the magnetic 
scattering factor can be expressed as 
$$
f_{mag}=
\bordermatrix{     & \sigma & \pi  \cr
               \sigma^{\prime} & 0 & m_c{\rm sin}\theta \cr
               \pi^{\prime} & m_c{\rm sin}\theta & 0 \cr
            }
$$
where 
$m_c$ is the components of the spins along the $c$-axis and 
$\theta$ is the scattering angle \cite{Hannon,JPHill}. 
In this case, pure $\sigma$ -$\pi^{\prime}$ and 
$\pi$ - $\sigma^{\prime}$ channels 
have the same intensity, which agrees very well 
with our experimental results and verifies the interpretation 
in terms of magnetic scattering. 

The emergence of the magnetic $L$ = 2/3 and 4/3 peaks agrees well with 
the powder neutron diffraction measurement at 2 T. 
In order to assign the shoulder peaks 
around $L$ = 7/6, 
the data were fitted by three components 
of $L$ = 7/6, 8/7 and 6/5 
as shown in Fig.~\ref{peak} (c). 
These results therefore directly reveal that a large number of magnetic
phases coexist in zero magnetic field and, in particular that 
the $\uparrow \uparrow \downarrow$ configuration 
is realized even at zero magnetic field. 
We observed the magnetic peaks 
of $L=n/6$ with $n=$ 4, 5, 6, 7, 8, and 9. 
However,the temperature dependence varies 
for different $n$ and the peaks with $n=5$ and 7 
show the shift from ICM to CM 
peak position and the others do not. 
This indicates that all the different peaks cannot be 
due to one magnetic modulation with $L=1/6$ 
but belong to different magnetic stacking sequences. 
This is also reflected in the observed field dependent behaviour. 

\begin{figure}
\begin{center}
\includegraphics[width=9cm]{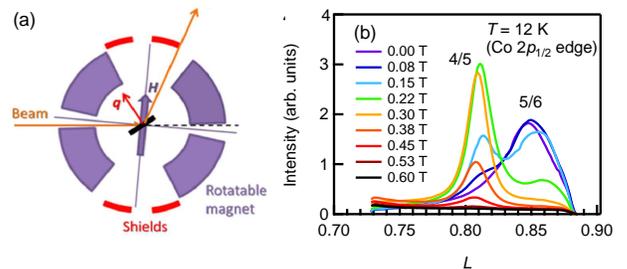}
\caption{(Color online) 
Geometry for RSXS under magnetic field (a), 
and magnetic peaks observed under various field (b).}
\label{geometry}
\end{center}
\end{figure}

The geometry of x-ray beam and the superconducting magnet 
for the field-dependent experiment is shown 
in Fig.~\ref{geometry} (a). 
Figure \ref{geometry} (b) shows the magnetic peaks at 12 K around $L$ = 4/5. 
The denoted values of the magnetic field are the $c$-axis component 
because the $ab$ component is irrelevant 
to the magnetic structures due to the strong 
Ising-like anisotropy \cite{ishiwataPRL,SaitoJMMM}. 
The $L$ = 5/6 peak has strong intensity around 
$H$ = 0 T, while the 4/5 peak is stabilized around $H$ = 0.2 T, 
indicating that 
the magnetic peaks with different $L$ behave differently 
under magnetic fields.  

\begin{figure}
\begin{center}
\includegraphics[width=9cm]{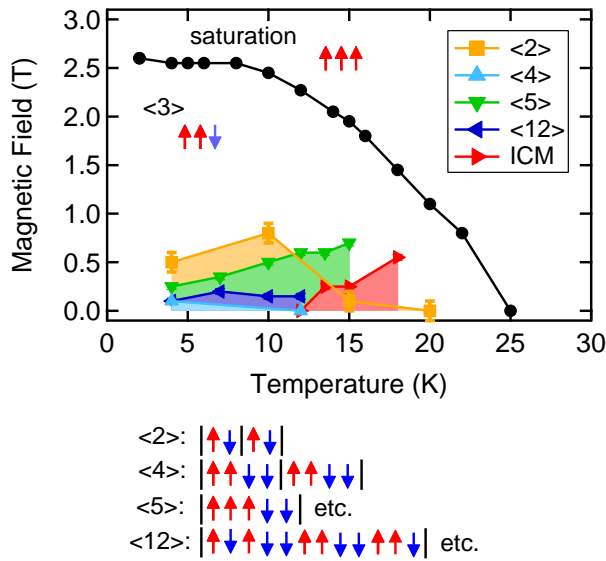}
\caption{(Color online) 
Magnetic phase diagram of SrCo$_6$O$_{11}$ determined by 
RSXS measurements. The phase 
boundary between $\uparrow\uparrow\uparrow$ and 
$\uparrow\uparrow\downarrow$ states was 
determined by magnetization measurements.}
\label{phase}
\end{center}
\end{figure}

In this way, we have been able to explore the entire $H$-$T$ 
diagram of this complex magnetic system 
as shown in Fig.~\ref{phase}. 
The phase boundary between $\uparrow\uparrow\uparrow$ and 
$\uparrow\uparrow\downarrow$ states was 
determined by magnetization measurements, and 
the other boundaries were determined by the present RSXS results. 
Here, $\braket{n}$ represents the magnetic periodicities. 
Since the SrCo$_6$O$_{11}$ unit cell contains two equivalent Co(3) 
Bragg planes along $c$, (002) is the first allowed 
structural reflections and (001) corresponds to a simple 
$\uparrow\downarrow\uparrow\downarrow$ 
antiferromagnetic order. Therefore 
$\braket{2}$ corresponds to $L=1$, 
$\braket{4}$ to $L=3/2$, 
$\braket{5}$ to $L=4/5$, and 
$\braket{12}$ to $L=5/6$. 
The phase diagram demonstrates that various magnetic orderings 
with different periodicities are formed in the low temperature and 
low field region. Obviously, the energies of these magnetic 
structures are quite close, and the corresponding energy differences 
sensitively depend on temperature and magnetic fields. 
A similar behavior has been observed in CeSb, 
which also has various magnetic orderings 
depending on temperature and field \cite{Bak,JRM1,JRM2}. 
This phenomenon is called ``the devil's 
staircase'', 
whose mechanism is described by the axial next nearest neighboring 
Ising (ANNNI) model \cite{Bak,JRM1,JRM2,Ohwada,JvB,Bak2,Selke,Nakanishi}. 
The ANNNI model describes competing interactions 
between nearest and next-nearest Ising spins, 
which yields various magnetic orderings with close energies. 

In SrCoO$_6$ the situation is very similar 
in that i) Co(3) has a strong Ising anisotropy and 
ii) our RSXS results reveal a coexistence of various essentially 
degenerate magnetic phases. 
We therefore conclude that the SrCoO$_6$ indeed exhibits a 
devil's-staircase scenario, 
i.e. a coexistence of a large number of magnetic periodicities 
with almost the same energies. 
Such coexistence is destroyed by the application of 
an external magnetic field, selecting only those 
phases which are energetically favorable now, 
leading to the magnetization plateaus observed 
in macroscopic measurements. Therefore, 
SrCo$_6$O$_{11}$ is considered as 
the first example of the devil's staircase 
which was found in a 3$d$-electron spin system. 

However, in order to explain ordered structures 
with long periodicities, 
one needs magnetic interactions that go well 
beyond the nearest neighbors. A plausible explanation 
can be provided by the Ruderman-Kittel-Kasuya-Yosida (RKKY) 
interaction via the metallic planes. 
SrCo$_6$O$_{11}$ has strong coupling between conduction electrons and 
localized spins, where RKKY interactions play the most important role. 
Consequently the very complex behaviour of the magnetically 
highly frustrated SrCo$_6$O$_{11}$ is far beyond the description 
of the simple ANNNI model. It may be better described 
by a more complex model with both localized spins 
and conduction electrons. 

Interestingly the observed 
very complex microscopic magnetic behavior 
seems to be also reflected 
in the macroscopic material properties: 
The out-of-plane magnetization 
at low temperatures in low magnetic fields 
in Fig.~\ref{fig1} (d) show hysteresis and 
additional plateaus around 1/5 and 1/6. 
These plateaus are 
caused by the phases of $L=4/5$ and 5/6, 
which have a smaller magnetization than the 1/3 phase. 
However, 
finally, the strongest ferrimagnetic phase ($L=4/3$) 
is the only remaining phase in higher magnetic fields, 
which creates the 1/3 magnetization plateau. 
This field- and temperature dependent magnetization 
is closely linked to the measured 
normalized out-of-plane resistivity $\rho_c(H)/\rho_c(0)$, 
which is characterized by several plateaus and hysteretic 
behaviour in the low-field region 
as shown in Fig.~\ref{fig1} (a). 
These macroscopic properties can be easily understood 
on the basis of the observed magnetic phase diagram: 
The magnetic phases in the low-field region are connected 
with anomalies in magnetization and resistivity 
since the magnetic field is able to select and stabilize 
single phases out of this ``nearly degenerate ground state''. 
Hence, the basic mechanism behind functionality of 
the novel spin valve system SrCo$_6$O$_{11}$ is a very high degree of magnetic 
frustration leading to a complex magnetic phase mixture 
whose delicate balance can be easily modified even 
by small magnetic fields.

\begin{figure}[H]
\begin{center}
\includegraphics[width=9cm]{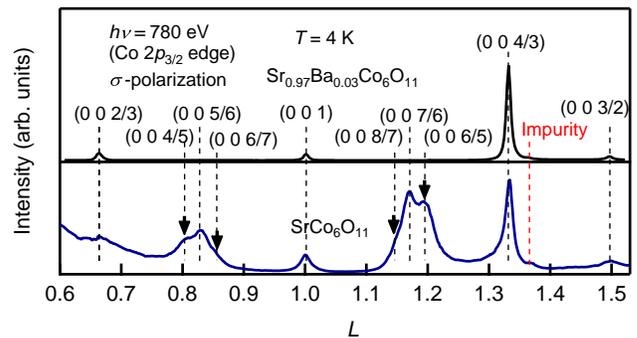}
\caption{(Color online) Magnetic peaks at zero field of 
SrCo$_6$O$_{11}$ and Sr$_{0.97}$Ba$_{0.03}$Co$_6$O$_{11}$ at 4 K.}
\label{phase1}
\end{center}
\end{figure}

Besides the related magneto-resistive 
functionality, frustrated magnets are 
also very sensitive to chemical doping. 
In contrast to robust systems, which require substantial 
doping to alter material properties, here a fine tuning 
of material properties should be possible by very low doping, 
in this way preventing unwanted side-effects. 
This is demonstrated in the present study 
by Ba substitution of Sr of only a few percent 
as shown in Fig.~\ref{phase1}. 
In Sr$_{0.97}$Ba$_{0.03}$Co$_6$O$_{11}$ one 
cannot see any magnetic peaks around $L=$ 5/6 or 7/6, 
demonstrating that 
small Ba substitution of only 3\% 
destroys almost the degenerate ground states, 
and consequently shifts the  
$1/3\rightarrow3/3$ functional step 
to lower magnetic fields. Stronger substitution of 10\% 
shifts this step 
to 0 T, i.e. the system has a FM ground state 
as shown in Fig.~\ref{fig1} (c). 

In summary, we have investigated the magnetic structures of 
SrCo$_6$O$_{11}$ bulk single crystals.  
We observed the first devil's staircase behavior in a $3d$ system 
where coupling is mediated by RKKY interaction. 
This is a consequence of a highly frustrated magnetic systems. 
The ground state, where there is a coexistence 
of various magnetic periodicities with almost 
the same energies, is very susceptible to 
magnetic fields and finally is the responsible 
mechanism behind the observed macroscopic functionality. 
In connection with the layered structure and Ising like anisotropy, 
this generates a spin valve functionality 
in a single phase material, 
which usually requires complex heterostructures. 
Furthermore, having frustration as the fundamental 
mechanism, this system can be easily tunable, 
rising the hope for engineered system properties, 
which has been demonstrated here by studying 
the behavior depending on very small amount of 
Ba doping. 

This work was supported by the Japan Society for the Promotion of Science (JSPS)
through the ``Funding Program for World-Leading Innovative R\&D on Science
and Technology (FIRST Program)'' 
initiated by the Council for Science and Technology Policy (CSTP) 
and in part by JSPS Grant-in-Aid for Scientific Research(S) 
No.~24224009. 
This work was also partially supported by the Ministry of Education, 
Culture, Sports, Science and Technology of Japan 
(X-ray Free Electron Laser Priority Strategy Program) and 
by a research granted from The Murata Science Foundation. 
S. P. and J. G. thank the DFG for the support through the Emmy 
Noether Program (Grant GE 1647/2-1).


\begin{thebibliography}{99}
\bibitem{nobel1}
M. N. Baibich, J. M. Broto, A. Fert, F. N. Van Dau, F. Petroff, 
P. Etienne, G. Creuzet, A. Friederich, and J. Chazelas, 
Phys. Rev. Lett. {\bf 61}, 2472 (1988).
\bibitem{nobel2}
G. Binasch, P. Gr\"{u}nberg, 
F. Saurenbach, and W. Zinn, 
Phys. Rev. B {\bf 39}, 4828 (1989).
\bibitem{RamirezMn}
A.~P. Ramirez, J. Phys.: Condens. Mat. {\bf 9},  8171  (1997).
\bibitem{RaoMn}
C.~N.~R. Rao, A. Arulraj, A.~K. Cheetham, and B. Raveau, 
J. Phys.: Condens. Mat. {\bf 12},  R83  (2000).
\bibitem{PrellierMn}
W. Prellier, P. Lecoeur, and B. Mercey, J. Phys.: 
Condens. Mat. {\bf 13},  R915 (2001).
\bibitem{Hungry}
A.~M. Haghiri-Gosnet and J.~P. Renard, J. Phys. D: 
Appl. Phys. {\bf 36},  R127 (2003).
\bibitem{TokuraMn}
Y. Tokura, Rep. Prog. Phys. {\bf 69},  797  (2006).
\bibitem{synthesis} 
S. Ishiwata, D. Wang, T. Saito, M. Takano, 
Chem. Mater. {\bf 17}, 2789 (2005).
\bibitem{ishiwataPRL} 
S. Ishiwata, I. Terasaki, F. Ishii, N. Nagaosa, H. Mukuda, Y. Kitaoka, 
T. Saito and M. Takano, Phys. Rev. Lett. {\bf 98}, 217201 (2007).
\bibitem{CCO}
H. Wu, M. W. Haverkort, Z. Hu, D. I. Khomskii, and L. H. Tjeng, 
Phys. Rev. Lett. {\bf 95}, 186401 (2005). 
\bibitem{SaitoJMMM} 
T. Saito, A. Williams, J. P. Attfield, T. Wuernisha, 
T. Kamiyama, S. Ishiwata, Y. Takeda, Y. Shimakawa and M. Takano, 
J. Magn. Magn. Mater. {\bf 310}, 1584 (2007).
\bibitem{Bak} 
P. Bak, Rep. Prog. Phys. {\bf 45}, 587 (1982).
\bibitem{JRM1} 
J. R.-Mignod, P. Burlet, J. Villain, H. Bartholin, 
Wang Tcheng-Si, D. Florence and O. Vogt, 
Phys. Rev. B {\bf 16}, 440 (1977).
\bibitem{JRM2} 
J. R.-Mignod, J. M. Effantin, 
P. Burlet, T. Chattopadhyay, L. P. Regnault, H. Bartholin, 
C. Vettier, O. Vogt, D. Ravot and J. C. Achart, 
J. Magn. Magn. Mater. {\bf 52}, 111 (1985).
\bibitem{Ohwada} 
K. Ohwada, Y. Fujii, N. Takesue, M. Isobe, Y. Ueda, H. Nakao, 
Y. Wakabayashi, Y. Murakami, K. Ito, Y. Amemiya, 
H. Fujihisa, K. Aoki, T. Shobu, Y. Noda, and N. Ikeda, 
Phys. Rev. Lett. {\bf 87}, 086402 (2001).
\bibitem{JvB} 
J. von Boehm and P. Bak, Phys. Rev. Lett. {\bf 42}, 122 (1979).
\bibitem{Bak2} 
P. Bak and J. von Boehm, Phys. Rev. B {\bf 21}, 5297 (1980).
\bibitem{Selke} 
W. Selke and P. M. Duxbury, Z. Phys. B: Cond. Matt. {\bf 57}, 49 (1984).
\bibitem{Nakanishi} 
K. Nakanishi, J. Phys. Soc. Jpn. {\bf 58}, 1296 (1989).
\bibitem{RSXDreview} 
J. Fink, E. Schierle, E. Weschke and J. Geck, 
Rep. Prog. Phys. {\bf 76}, 056502 (2013).
\bibitem{Wadati} 
H. Wadati, J. Okamoto, M. Garganourakis, V. Scagnoli, U. Staub, 
Y. Yamasaki, H. Nakao, Y. Murakami, M. Mochizuki, M. Nakamura, 
M. Kawasaki, and Y. Tokura, Phys. Rev. Lett. {\bf 108}, 047203 (2012). 
\bibitem{Partzsch} 
S. Partzsch, S. B. Wilkins, J. P. Hill, E. Schierle, E. Weschke, 
D. Souptel, B. B\"uchner, and J. Geck, 
Phys. Rev. Lett. {\bf 107}, 057201 (2011).
\bibitem{Zhou} 
S. Y. Zhou,Y. Zhu, M. C. Langner, Y.-D. Chuang, 
P.Yu, W. L.Yang, A. G. Cruz Gonzalez, 
N. Tahir, M. Rini, Y.-H. Chu, R. Ramesh, D.-H. Lee, Y. Tomioka, 
Y. Tokura, Z. Hussain, and R. W. Schoenlein, 
Phys. Rev. Lett. {\bf 106}, 186404 (2011).
\bibitem{Beale} 
T. A. W. Beale, S. B. Wilkins, R. D. Johnson, 
S. R. Bland, Y. Joly, T. R. Forrest, D. F.McMorrow, 
F. Yakhou, D. Prabhakaran, A. T. Boothroyd, and P. D. Hatton, 
Phys. Rev. Lett. {\bf 105}, 087203 (2010).
\bibitem{Smadici} 
S. Smadici, J. C. T. Lee, S. Wang, 
P. Abbamonte, G. Logvenov, A. Gozar, C. Deville Cavellin, 
and I. Bozovic, Phys. Rev. Lett. {\bf 102}, 107004 (2009).
\bibitem{Hannon} 
J. P. Hannon, G. T. Trammell, M. Blume, and D. Gibbs, 
Phys. Rev. Lett. {\bf 61}, 1245 (1988). 
\bibitem{JPHill} 
J. P. Hill and D. F. McMorrow, Acta Cryst. {\bf 52}, 236 (1996).
\end{thebibliography}
\end{document}